# Defect tolerant device geometries.

Basita Das[1], Zhifa Liu[1], Irene Aguilera[1], Uwe Rau[1] and Thomas Kirchartz[1,2]
[1]IEK5-Photovoltaik, Forschungszentrum Jülich, 52425 Jülich, Germany
[2]Faculty of Engineering and CENIDE, University of Duisburg-Essen, Carl-Benz-Str. 199, 47057 Duisburg, Germany

Abstract:

The term defect tolerance is widely used in literature to describe materials such as lead-halides which exhibit long non-radiative lifetimes of carriers despite possessing a large concentration of point defects. Studies on defect tolerance of materials mostly look at the properties of the host material and/or the chemical nature of defects that affect the capture coefficients of defects. However, the recombination activity of a defect is not only a function of its capture coefficients alone but are also dependent on the electrostatics and the design of the layer stack of a photovoltaic device. Here we study the influence of device geometry on defect tolerance by combining calculations of capture coefficients with device simulations. We derive generic device design principles which can inhibit recombination inside a photovoltaic device for a given set of capture coefficients based on the idea of slowing down the slower of the two processes (electron and hole capture) even further by modifying electron and hole injection into the absorber layer. We use the material parameters and typical p-i-n device geometry representing methylammonium lead halide perovskites solar cells to illustrate the application of our generic design principles to improve specific devices .

## I. Introduction

With the recent interest in metal-halide perovskites for photovoltaic and optoelectronic applications, the term defect tolerance[1–8] has become omnipresent in the scientific literature. While it has been coined in the context of computational material screening[6], lead-halide perovskites were the first incarnation of a defect tolerant semiconductor with an antibonding valence band that would work particularly well in solar cells and light emitting diodes[5,7]. One of the key reasons for this success was that solution processed polycrystalline layers of lead-halide perovskites in a wide range of stoichiometries and with varying cations and halides showed long charge carrier lifetimes[9], high luminescence quantum yields[10–14] and subsequently high open-circuit voltages[14–17] and efficiencies if used as a solar cell. In most classical semiconductors, deep defects form at grain boundaries, surfaces, interfaces and in the bulk of the film and accelerate recombination. The existence of these defects makes it usually quite difficult to achieve good electronic properties without investing in complicated and (usually) energy intensive fabrication techniques such as single crystal growth and epitaxy. In halide perovskites, the impact of defects is less severe, because most intrinsic point defects create rather shallow energy levels[18] making non-radiative transitions less likely than in most other semiconductors. The prevalence of shallow defects is caused by the rather uncommon antibonding valence band that causes most atomic orbitals to be in the conduction or valence band of the crystal. Thus, when bonds are broken (e.g.

in the presence of a vacancy), the resulting orbitals are mostly inside the bands or close to the band edges rather than in the middle of the band gap as is the case for most covalent semiconductors. While, the defect tolerance of a semiconductor is certainly an extremely important topic for understanding the success of lead-halide perovskites, so far there have been few attempts to discuss how the device geometry of a solar cell can affect the vulnerability of the device to defects.

Whether a defect causes substantial recombination not only depends on the density and energetic position of defects but also on the availability of electrons and holes to be captured by these defects. If for instance the material has an abundance of defects that preferentially trap holes, the capture of electrons is likely the rate limiting step. If the device geometry is then designed such that there is an excess of holes in the absorber layer of the device, electron capture would be further reduced, and the total device efficiency may go up. However, changing the properties of the device geometry has a multitude of consequences given that the device geometry has to allow efficient charge extraction, therefore provide selectivity at the different contacts, allow sufficient light absorption and efficient current flow minimizing resistive losses. Therefore, the perfect design of the device geometry under the explicit consideration of the properties of the dominant defects is a non-trivial task.

Here, we investigate the effect of device geometry on the recombination efficiency of the defects, and hence on the efficiency of the solar cell. As opposed to the popular assumption that defects have symmetric capture coefficients used in solar cell device simualtion[19,20], defects mostly have asymmetric capture coefficients, i.e. a defect is not as likely to capture an electron as a hole or vice-versa. From a device point of view, the net impact of defects can be reduced by modulating the carrier concentration inside the device such that one of the capture rates is substantially slower than the other. Unfortunately, there is no one device geometry that will improve performance of all devices. However, the current work offers certain generic design principles which when implemented after identifying the dominant recombination levels will help reduce the recombination through the device.

This work explains how asymmetric device architectures improves open-circuit voltage of a device as seen from real device data[14] by mitigating defect mediated recombination inside the device. It also discusses the role of the built-in-voltage in controlling the carrier concentration inside the device which then directly affects the solar cell device performance[21,22]. We highlight the importance of the electrostatic potential drop across the device[23] and how it alters the efficiency of the device by either enhancing or mitigating the recombination through defects. We also show that the perovskite solar cell efficiency is limited by slow charge extraction due to the low mobility organic transport layers[24] and can be improved if the low mobility transport layers were to be replaced with high mobility transport layers. Besides, this work also considers the importance of atomistic calculations[25–29] or generalized models[30,31] in determining realistic values of capture coefficients and how these parameters can improve solar cell device simulation to help it go beyond the current state-of-the-art of symmetric capture coefficients[19,20,24].

## II. Theoretical Background

### A. Defect-mediated recombination

Defects occurring within the bandgap of a semiconductor enhance recombination in the device as they form an alternative recombination channel for the charge carriers besides band to band recombination. The theoretical description of this accelerated recombination via defects is based on work by Shockley, Read and Hall[32,33] and is therefore known as Shockley-Read-Hall (SRH) recombination. To help understanding the results and discussion in the remainder of the article, we will briefly introduce the key terminology of SRH-recombination in the following.

Recombination via singly charged defects (donor-like or acceptor-like) occurs when the defect goes through a cycle constituting two consecutive processes of electron capture and hole capture. The rate of recombination via such a defect depends on the electron capture rate (per unit time) $nk_n$ [s$^{-1}$] and hole capture rate $pk_p$ [s$^{-1}$], where $n$ and $p$ are the electron and hole concentration, and $k_n$ [cm³/s] and $k_p$ [cm³/s] are the electron and hole capture coefficient, respectively. The rate of recombination (per unit time and volume) is then given as

$$R_{\text{SRH}} = N_T \eta_R, \tag{1}$$

where $N_T$ [cm$^{-3}$] is the defect density of the semiconductor and the recombination efficiency $\eta_R$ [s$^{-1}$] of a defect level *i.e.* the number of recombination events occurring at a defect per unit time is given as

$$\eta_R = \frac{k_n k_p (np - n_0 p_0)}{nk_n + pk_p + e_n + e_p}. \tag{2}$$

Furthermore, for defect levels lying well inside the two quasi-Fermi levels the emission coefficients are such that $e_n \gg nk_n$ and $e_p \gg pk_p$ and by assuming $np \gg n_0 p_0$, the recombination efficiency can be simplified as follows

$$\eta_R \approx \frac{nk_n pk_p}{nk_n + pk_p}. \tag{3}$$

Inside a device, the electron and hole carrier concentrations $n_0$ and $p_0$ in equilibrium, and $n$ and $p$ in typical operating conditions, respectively, are affected by the workfunctions and doping densities of the different layers in the device whereas the capture coefficients $k_{n/p}$ are determined by the chemical nature and energetic position of a defect within the bandgap. The electron and hole emission coefficients $e_n$ and $e_p$ depend on the respective capture coefficients and the position of the quasi-Fermi level splitting given by the typical operating conditions of the device. Recombination efficiency is therefore affected by the combined properties of the device geometry and the nature of the defects.

However, SRH statistics is agnostic about the origin of the capture coefficients and makes no theoretical prediction of the values representative of $k_{n/p}$ in materials. In absence of a model to predict the values of capture coefficients, state-of-the-art device simulation of photovoltaics uses capture coefficients obtained from experimental methods[20] or heuristic assumptions[34] to calculate electron hole capture lifetimes $\tau_{n/p} = (N_T k_{n/p})^{-1}$. The frequently used assumption of symmetric capture coefficients $k_n = k_p$ for all defect levels within the bandgap in a device operating in a high injection scenario $n = p$ leads to a simplified picture where all defects within the two quasi-Fermi are equally recombination active. This behavior of defects is often used in textbooks for photovoltaics to illustrate the impact of capture and emission coefficients on the SRH recombination rate or recombination lifetime[35,36]. However, equal or at least similar values of the capture coefficients are not necessarily a typical or even a likely scenario[30,37].

To go beyond the state-of-the-art device simulation in photovoltaics involving symmetric capture coefficients, we either need more experimental data on capture coefficients of novel materials or, alternatively, models to calculate trap energy dependent capture coefficients. These can then be used in combination with SRH statistics to get a trap energy dependent recombination efficiency and SRH lifetime.

## B. Capture coefficients

Recently we presented a generalized microscopic model[30,31] based on the harmonic oscillator approximation to calculate multiphonon capture coefficients using material properties of the host semiconductor and energetic position of the trap within the bandgap of the material. The model, adapted from the work of Markvart[38,39] and Ridley[40–42], uses the so called 'quantum defect model' to model the defects and characterizes their degree of localization by the quantum defect parameter, $\upsilon_T$. The parameter $\upsilon_T$, defined as the square root of the depth $\Delta E_{min}$ of any defect level $E_T$ from the nearest band w.r.t the depth of a shallow defect given by effective Rydberg energy $R_H$ ($\upsilon_T = \sqrt{R_{H^*}/\Delta E_{min}}$), ensures that the higher the depth of the defect, the more localized it is. The quantum defect parameter $\upsilon_T$ diminishes as the depth of the defect $\Delta E_{min}$ form the nearest band increases and assumes the smallest value when the defect is at midgap. The Huang-Rhys factor $S_{HR}$ measuring the displacement of the lattice from its mean position in the vicinity of the defect is an inverse function of $\upsilon_T$ and increases with increasing $\Delta E_{min}$. Besides, the type of coupling (polar optical coupling for polar semiconductors or optical deformation potential coupling for elemental semiconductors or less polar semiconductors) also the charge state of the defect, $\mu$, is considered in the calculation of $S_{HR}$. The Huang-Rhys factor is then used to calculate the capture coefficients of defects (see Table I in supplementary information(SI) for expressions)[30] in the material. To consider the Coulomb attraction between an oppositely charged defect state and a captured carrier the Sommerfeld factor $s_a$ is multiplied with the electron capture coefficient $k_n$ of a donor-like defect and with the hole capture coefficients $k_p$ of an acceptor-like defect. When the trap energy dependent capture coefficients are used as inputs to the SRH theory, the energy dependence of the recombination efficiency is reflected.

It is evident from the expression of the capture coefficients given in Table I of the SI that $k_{n/p}$ are dependent on the depth $\Delta E$ of a trap. The depth $\Delta E$ of the trap is always measured with respect to the bandedge from which the carrier is captured as shown in Fig. 1a. This ensures that hole capture coefficient $k_p$ decreases rapidly as the defect level moves away from the valence bandedge $E_V$ whereas the electron capture coefficient $k_n$ decreases as the defect moves away from the conduction bandedge $E_C$ as shown in Fig. 1b. The faster than exponential decay of capture coefficients w.r.t to $\Delta E$ measured from the respective bands of transition makes $k_n$ and $k_p$ highly asymmetric except for around midgap where the two are comparable or equal. The generalized model for capture coefficients predicts that in methylammonium lead iodide (MAPI) perovskite even defects lying within 200 meV on either side of midgap have $k_n$ and $k_p$ values differing from each other by $10^5$ to $10^2$ orders of magnitude as shown in Fig. 1b. This idea of asymmetric capture coefficients[37] contrasts with the usual assumption of symmetric capture coefficients[20,34] for any defect level and has a huge impact on the recombination efficiency of a device.

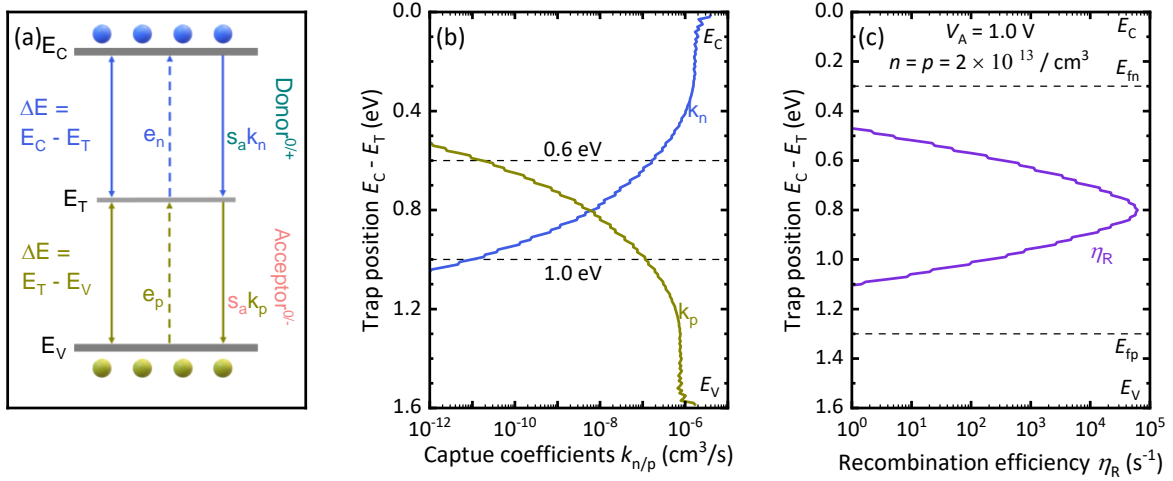

Fig. 1 Trap energy dependent capture coefficients and recombination energy. (a) Schematic indicating the depth of trap $\Delta E = E_C - E_T$ for transitions taking place from conduction band or $\Delta E = E_T - E_V$ for transitions taking place from valence band. The electron capture coefficient $k_n$ and emission coefficient $e_n$, hole capture coefficient $k_p$ and emission coefficient $e_p$. The $k_n$ of a donor-like trap and $k_p$ of an acceptor-like trap are multiplied with Sommerfeld factor $s_a$ to account for Coulomb interaction between oppositely charged defect state and captured carrier. (b) The capture coefficients $k_{n/p}$ (cm³/s) as functions of trap position measured from the edge of the conduction band. The capture rates are obtained by multiplying the electron and hole carrier concentration with the electron and hole capture coefficients. The carrier concentration is considered to be equal and calculated at an applied voltage $V_A = 1.0$ V. (c) The recombination efficiency $\eta_R$ (s⁻¹) as a function of trap position $E_C - E_T$.

In a simple scenario of high injection of carriers $n = p$, the capture rates $nk_n$ and $pk_p$ follow the same decaying nature of the respective coefficient w.r.t $\Delta E$ measured from bandedge of transition. Inserting these rates into Eq. 2 leads to the result shown in Fig. 1c, where within the two quasi-Fermi level splitting set by the applied voltage $V_A$, $\eta_R$ is given by the slower of the two rates except for midgap where the two rates are symmetric and a combination of the two rates gives $\eta_R$. We use this parameter $\eta_R$ giving the number of recombination event at a defect per unit time to characterize a "deep" defect. Our way of classifying a "deep" defect is based on the defect's recombination efficiency and not merely on its energetic position inside the bandgap. Accordingly, a defect is classified as "deep" when the defect attains its most detrimental state. And the most detrimental state is attained when as a result of the $n$ and $p$ values determined by the operating condition, the defect satisfies the two criteria

$$nk_n = pk_p \tag{i}$$

$$nk_\text{n} \gg e_\text{n} \text{ and } pk_\text{p} \gg e_\text{p}. \tag{ii}$$

The second criterion is valid for defects occurring inside the two quasi-Fermi level of the device and the first criterion then determines the *n* and *p* values which makes the electron and hole capture rate equal and thus leading a defect to it most detrimental state. In our simple scenario of *n* = *p* the first criterion is satisfied only at midgap thus making a midgap defect the most recombination active or "deep" defect.

However, the observation that only midgap defects are "deep" is a result of equal electron and hole concentration and will change as soon as we consider asymmetric carrier concentrations inside a device. It is crucial to realize that even though the capture coefficients are intrinsic properties of the defect, the recombination activity is not. The recombination efficiency of a defect is determined by the *n* and *p* values and hence can change with the change in carrier concentration. Inside a device the same defect level can vary in recombination activity depending upon its position within the device. The significance of these two criteria is that when satisfied it indicates an occupation probability of the defect $f_\text{T} \sim 0.5$, and maximum recombination efficiency for the defect. When the second criterion is still valid, but *n* and *p* values are such that $nk_\text{n} \neq pk_\text{p}$, the recombination efficiency of the particular defect is limited by the slower of the two rates. From a device perspective, the defects satisfying the two criteria are also likely to be the most detrimental of all defect levels in the device, at that particular working condition. The energy and position dependence of $\eta_\text{R}$ inside a device elucidates that not all defects are equally detrimental to device performance and whether a defect is "deep" or not is determined by the working condition of the device.

To understand the realistic impact of asymmetric capture coefficients of defects inside a device, we need to go beyond the simple scenario of *n* = *p* and use asymmetric capture coefficients obtained from our general model based on the harmonic oscillator approximation in combination with a photovoltaic device simulator to map the recombination efficiency of trap levels as a function of both trap depth *ΔE* and position inside the device.

## III. Results and Discussion

The criteria for "deep" defects discussed in the previous section are subject to the harmonic oscillator approximation. However, within this approximation, they are also valid in situations where $n \neq p$ and $k_\text{n} \neq k_\text{p}$. Thus, any defect satisfying the second criterion and having asymmetric capture coefficients can act as highly recombination active site when $nk_\text{n} = pk_\text{p}$. In Fig. 2, we use the capture coefficients calculated in Fig. 1b using the harmonic approximation for MAPI to map the recombination efficiency as a function of $n/p$ and $k_\text{n}/k_\text{p}$. The key insights we gain are:

i. The further below the defect is from midgap, the higher the electron concentration should be to obtain $nk_\text{n} = pk_\text{p}$. Alternatively, when a defect is more likely to capture a hole, $k_\text{n} \ll k_\text{p}$; the lower the electron concentration *n* is compared to the hole concentration *p* ($n \ll p$), the lower the recombination efficiency $\eta_\text{R}$ will be.

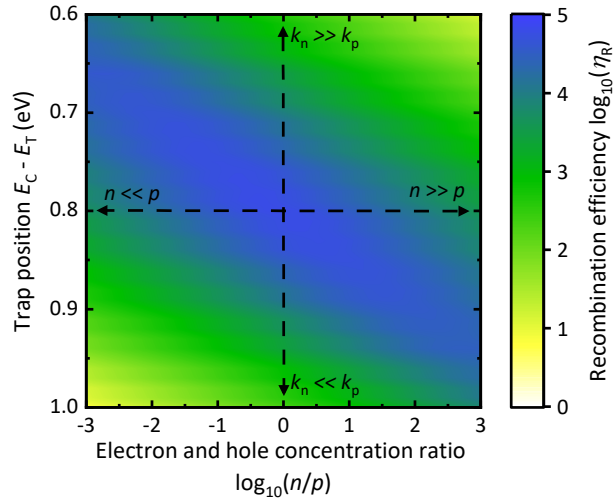

Fig 2 The variation of recombination efficiency $\eta_R$ as function of realtive values of electron and hole concentration ($n/p$) and the position of the trap. Recombination efficiency increases when the slower of the two capture rates increases in magnitude. For a particualar defect level with fixed capture coefficients this is determined by the carrier densities. When the defect is below the midgap level and closer to the valence band, $\eta_R$ increases from left to right with increase of $n$ with respect to $p$. When the defect is above the midgap level and close to the conduction band, the trend is reversed and $\eta_R$ increases from right to left with increase of $p$ with respect to $n$. The closer the defect is to the midgap level, the more symmetric are coefficients and $\eta_R$ peaks for more symmetric values of $n$ and $p$.

ii. The further above the defect is from midgap, the higher the hole concentration should be to obtain $nk_n = pk_p$. Alternatively, when a defect is more likely to capture an electron, $k_n \gg k_p$; the lower the hole concentration $p$ is compared to the electron concentration $n$ ($n \gg p$), the lower the recombination efficiency $\eta_R$ will be.
iii. The closer the defect is to midgap, the more comparable the carrier concentration should be to obtain $nk_n = pk_p$. Alternatively, when a defect is equally likely to capture a hole or an electron, $k_n \approx k_p$; either $n \gg p$ or $n \ll p$ will obtain a lower recombination efficiency $\eta_R$.

Hence, while dealing with asymmetric capture coefficients, it is better to have asymmetric carrier concentrations which keeps the rates of capture vastly different from each other and prevents the situation of $nk_n = pk_p$.

The insights gained so far are general and can act as guidelines to investigate different device geometries of different materials. A particularly interesting test-case to study is the class of lead-halide perovskite solar cells. Here, the absorber is fairly intrinsic and the ratio $n/p$ throughout the absorber depends to a large degree on properties of the electron and hole transport layers. For instance, the low permittivity of organic electron or hole transport layers, may cause a large drop of the electrostatic potential over the transport layers causing a smaller drop of electrostatic potential over the absorber layer. While this has been noted in the literature, so far, the impact of device electrostatics on recombination rates has not been discussed in much detail.

A perovskite solar cell (PSC) typically consists of an active layer sandwiched between an electron transport layer (ETL) and a hole transport layers (HTL) before the cathode and anode layer, respectively. In an ideal scenario, the ETL transports the electrons from the device to the cathode while blocking the holes and the HTL transports the holes from the active layer to the anode while blocking the electrons. As a result of the selectivity achieved by employing materials with different electron affinity, the mismatch of the workfunction of the HTL and ETL leads to a built-in-voltage $V_{bi}$ across the device. At an optimum working condition, the electrostatic potential drop across the device is the net difference between the built-in-voltage $V_{bi}$ and the applied voltage $V_A$ and the electrostatic potential then determines the electron and hole concentration inside the device.

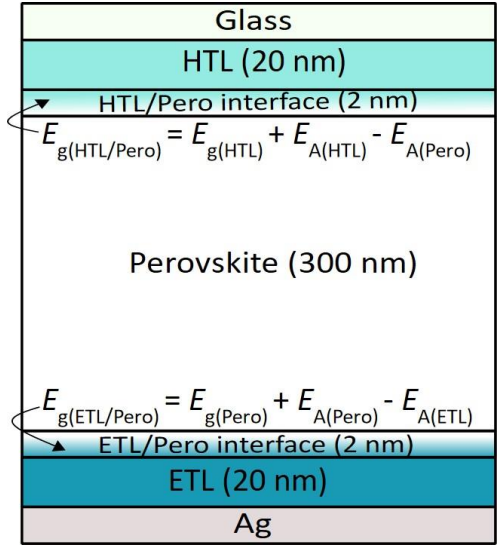

Fig. 3 P-i-n device structure of the perovskite solar cell under study. The generic hole transport layer (HTL) and electron transport layer (ETL) are symmetric differing only in their electron affinity ($E_A$). The interface layers are characterized by parameters of the absorber except for their bandgaps are given according to expressions shown in the figure.

We employ a device structure, with a perovskite absorber layer of 300 nm sandwiched between two generic charge transport layers each of 20 nm thickness as shown in Fig. 3. The ETL and HTL are symmetric differing only in their value of electron affinity $E_A$. The HTL/Perovskite and ETL/perovskite interfaces are modelled by additional layers each of 2nm and characterized by material parameters of the absorber layer. However, the bandgap of this interface layer is given by a combination of the bandgap and the electron affinity of the two materials forming the interface as shown in Fig. 3. We also assume a constant trap density $N_T = 10^{15}$ / cm$^3$ across the absorber and interface layers (see Table II in SI for material parameters). We then perform certain thought experiments realizing different $n/p$ ratios through our generic device by changing the electrostatic potential drop across the absorber and the contacts to correlate those variations in the ratio $n(x)/p(x)$ to the efficiency of the solar cell.

## A. Symmetric device with varying built-in-voltage:

The built-in-potential is determined by the energetic mismatch between the workfunctions of the anode and the cathode. When the Schottky barrier height at both the contacts are zero, the built-in-potential is equal to the bandgap of the absorber. When the workfunctions of both the metal contacts are such that there is a non-zero Schottky barrier at the ETL side and as well as the HTL side, then the built-in-potential is $qV_{bi} = E_{g(abs)} - \xi_p - \xi_n$, where $\xi_p$ and $\xi_n$ are the Schottky barrier heights at the HTL / anode interface and the ETL / cathode interface, respectively. We consider two scenarios such that in one $\xi_p = \xi_n = 0$ eV; $qV_{bi} = 1.6$ eV and in another $\xi_p = \xi_n = 0.2$ eV; $qV_{bi} = 1.2$ eV. In our symmetric device, the hole concentration $p$ and the electron concentration $n$ vary such that $p \gg n$ towards the HTL side and $n \gg p$ towards the ETL side leading to a continuously varying $n / p$ ratio inside the device.

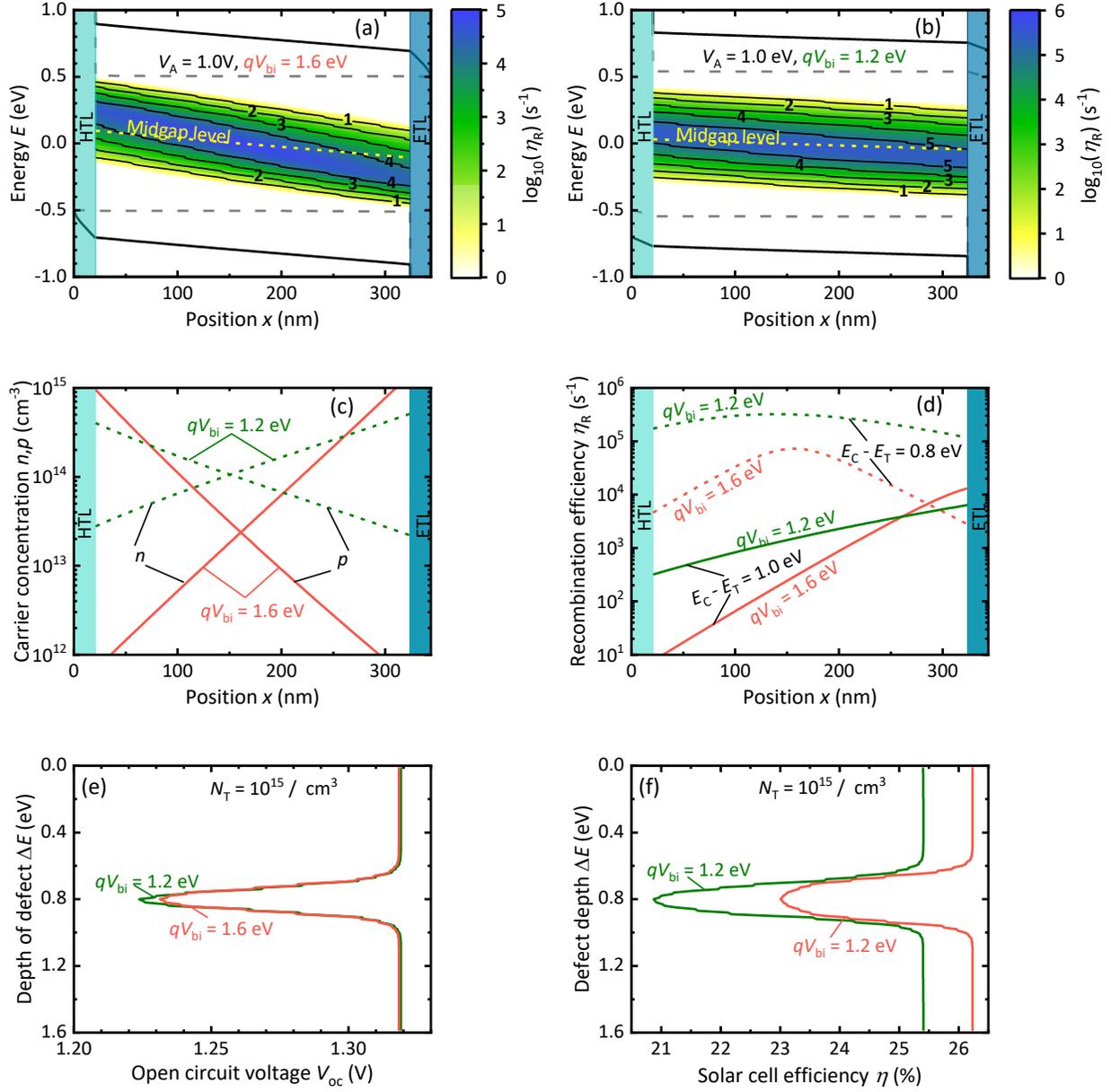

Fig. 4 The effect of changing the built-in-voltage $V_{bi}$ across the device on SRH recombination and subsequently on the device efficiency. (a) Recombination efficiency $\eta_R$ as a function of position inside a device and the energy level ($E_C - E_T$) of the trap calculated from the conduction band when $qV_{bi}$ = 1.6eV. This data is superimposed on the band diagram of the device. (b) Same as before but for $qV_{bi}$ = 1.2 eV. (c) The electron and hole concentration inside the device when $qV_{bi}$ = 1.6 eV and $qV_{bi}$ = 1.2 eV. (d) The SRH recombination rate $R_{SRH}$ of defects 1.0 eV and 0.8 eV away from the conduction band for $qV_{bi}$ = 1.6 eV and $qV_{bi}$ = 1.2 eV. Also, the direct recombination rate $R_{direct}$ for the two different built-in-voltages. (e) Open circuit voltage $V_{oc}$ plotted as a function of the position of the defect from the conduction band for $qV_{bi}$ = 1.6 eV and $qV_{bi}$ = 1.2 eV. (f) Solar cell efficiency $\eta$ (%) plotted as a function of the position of the defect from the conduction band for $qV_{bi}$ = 1.6 eV and $qV_{bi}$ = 1.2 eV

In Fig. 4a-b we plotted the band diagram of the two devices and then superimposed on it the logarithm of the absolute value of recombination efficiency $\eta_R$ as a function of both the position inside the device as well as the position of the defect as measured from the conduction bandedge. To explain the effect of the change in $qV_{bi}$ on the recombination efficiency we also plot the electron and hole concentration through the device at $qV_{bi}$ = 1.6 eV (solid red curves) and at $qV_{bi}$ = 1.2 eV (dashed green curve) in Fig. 4c. When $qV_{bi}$ = 1.6 eV, the hole to electron concentration ratio varies in range $p/n \approx 10^4$ to $10^{-4}$ from the HTL side to the ETL side as shown in Fig.4c. This suggest that any defect which has $k_n/k_p \approx 10^4$ has the potential to act as a "deep" defect at the HTL side and so does a defect with $k_n/k_p \approx 10^{-4}$ at the ETL side by satisfying the condition $nk_n = pk_p$. In case of MAPI as plotted in Fig. 1b, one such defect is about 0.6 eV from the conduction band with $k_n/k_p \approx 10^4$ whereas another one is about 1.0 eV away from the conduction band with $k_n/k_p \approx 10^{-4}$. Hence these two defects even with highly asymmetric capture coefficients satisfy the "deep" defect criteria of $nk_n = pk_p$ in different regions of the device. As we go inside the device from the HTL side the hole to electron concentration ratio decreases such that $10^4 > p/n > 10^{-4}$. This implies that as we move inside the device from the HTL side to the ETL side, any defect within 200 meV on either side of midgap, having capture coefficients such that $10^4 > k_n/k_p > 10^{-4}$, can potentially be turned into a "deep" defect or at least highly recombination active defect when the criteria $k_n/k_p \approx p/n$ is satisfied at a position inside the device. This highly recombination active region appears in the shade of blue in Fig. 4a and $\eta_R$ peaks around midgap at an approximate value of $10^5$ s$^{-1}$.

In comparison to $qV_{bi}$ = 1.6 eV, when non-zero Schottky barriers are introduced, the electron quasi-Fermi level move $\xi_n$ energy away from the conduction band at the ETL / metal interface and the hole quasi-Fermi level moves $\xi_p$ away from the valence band at the HTL / metal interface. This leads to reduction of the electron concentration towards the ETL and hole concentration towards the HTL. However, the change in $qV_{bi}$ increases carrier concentration everywhere else inside the device as well as the minority carrier concentration towards the selective contacts (holes towards ETL and electrons towards HTL). The change in both absolute value as well as relative value of electron and hole concentration has two effects on the recombination activity of the device.

In Fig. 4c when $qV_{bi}$ = 1.2 eV, the carrier concentration varies such as $p/n \approx 10$ to $10^{-1}$ from the HTL side to the ETL side, which implies that only defects with coefficients such that $k_n/k_p \approx 10$ to $10^{-1}$ are potential "deep" defects or the most detrimental of all the defects inside the device. Thus, according to the coefficients determined from harmonic oscillator approximation for MAPI (see Fig. 1b), defects within a 50 meV energy range on either side of midgap can be turned into "deep" defects by the electron and hole concentration at a particular position inside the device when $qV_{bi}$ = 1.2 eV. The narrowing down of the range of "deep" defects on reducing the built-in-voltage is related to the change of the relative values of electron and hole concentration inside the device. However, even though the range of "deep" defects narrowed down, comparing Fig. 4a-b reveals that the recombination efficiency of all defects increased and $\eta_R \approx 10^6$ s$^{-1}$ around midgap as a result of the increase in the absolute values electron and hole concentration as shown in Fig. 4c. As a result of reducing the built-in-voltage the recombination efficiency of below midgap defects ($k_n \ll k_p$) at every position inside the device increases by the amount of increase in electron concentration. The same is true for defects above midgap ($k_n \gg k_p$) but in this case, the increase in recombination efficiency is determined by the increase in the value of hole concentration.

In Fig. 4d we show the effect of the change $qV_{bi}$ and hence carrier concentration more explicitly by choosing to look at the recombination efficiency for two specific defect levels in MAPI with capture coefficients determined from the harmonic oscillator approximation. The defect level 1.0 eV away from the conduction band have asymmetric capture coefficients ($k_n \approx 10^{-11}$ cm$^3$/s and $k_p \approx 10^{-7}$ cm$^3$/s) and the one 0.8 eV away from the conduction band have symmetric capture coefficients ($k_n \approx k_p \approx 5e^{-9}$ cm$^3$/s). Both defect levels have a constant defect density of $N_T = 10^{15}$ cm$^{-3}$ across the active and the interfaces layers. The increase in electron concentration $n$ upon decreasing $qV_{bi}$ increased the rate of electron capture and hence $\eta_R$ of the defect 1.0 eV away from midgap through most of the device. However, the case of the defect at midgap with symmetric values of capture coefficient is bit more complicated. When $qV_{bi}$ = 1.6eV, $n$ and $p$ are vastly different, making the electron and hole capture rates very different even though the coefficients are symmetric and thus $\eta_R$ is determined by the slower of the two rates through most of the device. On the contrary when $qV_{bi}$ = 1.2 eV, the comparable $n$ and $p$ values yield very similar electron and hole capture rates having higher values compared to the capture rates at $qV_{bi}$ = 1.6 eV and thus $\eta_R$ is determined by a combination of both the capture rates.

In Fig.4e-f we show the overall effect of the change in built-in-voltage on the open-circuit voltage $V_{oc}$ and efficiency of the device $\eta$ (%). When the SRH recombination rate exceed the direct recombination rate $R_{SRH} > R_{direct}$, $V_{oc}$ and efficiency $\eta$ (%) decrease from their respective values given by the radiative limit. In the two devices discussed above this criterion is satisfied both by defects at midgap with symmetric coefficients as well as defects away from midgap with asymmetric capture coefficients. This deviation from the radiative limit within 200 meV on either side of midgap resulting in drop of $V_{oc}$ and $\eta$ (%) in this range. The drop in the values of $V_{oc}$ and $\eta$ (%) is higher when $qV_{bi}$ = 1.2 eV compared to when $qV_{bi}$ = 1.6 eV because of the higher $R_{SRH}$ in the low $qV_{bi}$ case. Even though the range of "deep" defects shrunk with the decrease in $qV_{bi}$ the actual overall recombination increased through the device due to the increase in the electron and hole concentration through the device. However, it should be noted that the direct recombination rate $R_{direct}$ increases substantially due to the increase in carrier concentration inside the absorber when $qV_{bi}$ decreases from 1.6 eV to 1.2 eV and the increased radiative recombination leads to the drop in the solar cell efficiency $\eta$ (%) beyond the 200 meV range on either side of midgap.

Thus, it is evident from the discussion above, that a built-in-voltage that helps in achieving relatively lower values of $n$ and $p$ inside the device also leads to relatively lower magnitudes of defect-mediated recombination as well as direct recombination. The rule is that if changing the built-in-potential $qV_{bi}$ leads to the decrease in even one of the carrier concentrations, then the capture rate of that carrier by all defect levels will also decrease. And if the capture rate that decreases due to decrease in carrier is the recombination limiting rate, then the recombination efficiency of the defect levels will decrease and as a result the efficiency of the device would increase. The results discussed above were for defect levels in the bulk. Devices limited by surface recombination shows similar trends as shown in Fig. S1 of the supplementary information. It is obvious that it is better to have vastly different electron and hole concentration inside the device so that the recombination rate can be slowed down by inhibiting one of the capture processes. However, even when $qV_{bi}$ is high, defects within 200 meV on either side of midgap act as deep defects. This will be particularly severe from a device perspective. If a device has higher concentrations of defects around midgap, this would result in very high levels of defect mediated recombination. Also, in the cases we studied, due to continuously varying electron and hole concentration, the recombination efficiency of every defect level varies continuously inside the device and so does the position of "deep" defects both in energy as well as position inside the device. To circumvent such scenarios, we study a device geometry such that $n \ll p$ or $p \ll n$ throughout the device.

### B. Asymmetric devices

To test our hypothesis about asymmetric carrier concentrations throughout the device we chose a device architecture with flat bands across the absorber layer. Such a device with symmetric ETL and HTL layer gives $n = p$ across the absorber and the interface layers and is expected to be highly detrimental for devices with defects at or around midgap. However, the recombination activity of defects close to midgap can be improved if we dope one of the contact layers as it would result into $n \gg p$ or $p \gg n$ across the absorber and interface layers depending on whether we dope the ETL layer with donor dopants or the HTL layer with acceptor dopants, respectively. If we want to induce an asymmetry of electron and hole concentrations in an intrinsic semiconductor by choosing asymmetric properties of contacts, we have to rely on the contacts being able to connect the majority carrier quasi-Fermi level at the contact with the one at the edge of the absorber without any substantial gradient in quasi-Fermi level. This requires either low thicknesses or sufficiently high mobilities in the undoped electron or hole transport layer. Because this cannot be taken for granted, we will first discuss the scenario where the contact layer mobilities are sufficiently high to minimize any voltage drops. In a second step (see section III.D), we discuss the opposite scenario where these mobilities are low enough to cause a substantial voltage drop.

In Fig. 5a-b we plot the logarithm of the absolute value of recombination efficiency of a symmetric and asymmetric flat-band device, respectively, superimposed on the band diagram of the respective device. In the symmetric device, the electron and hole carrier concentrations are equal and constant $n = p$ (as indicated by the position of the two quasi-Fermi level splitting from their respective bands) between the two interface layers resulting in the recombination efficiency peaking around midgap and decreasing rapidly beyond that as shown by the blue shaded region of Fig. 5a. In the asymmetric device show in Fig 5b, the ETL layer has a donor dopant concentration of $N_{d(ETL)} = N_{C(ETL)}$ and results in $n \gg p$ between the two interfaces. This asymmetric carrier concentration results in reduced recombination efficiency for all defects above midgap and within 100 meV below midgap, while increasing the same for defects which are about 100 meV – 300 meV below midgap. This is so because at midgap where $k_n \approx k_p$, $n \gg p$ makes $nk_n \gg pk_p$ and as a result $\eta_R$ is limited by the slower hole capture rate $pk_p$ and thus the defects around midgap do not anymore satisfy the "deep" defect criterion of $nk_n = pk_p$. However, the "deep" defect criterion is now satisfied by defects which have capture coefficients such that $k_n / k_p \approx p / n$, which in this case are defects situated about 100 – 300 meV below midgap.

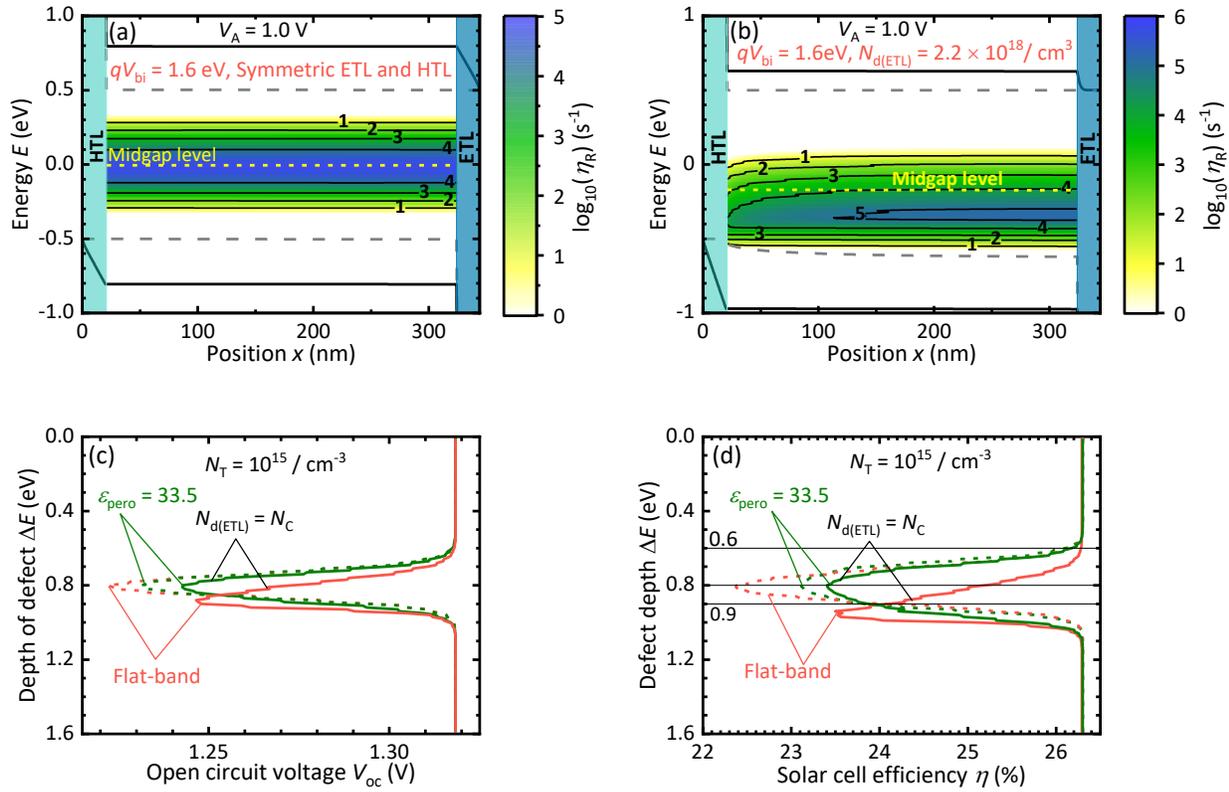

Fig 5 The Effect of constant but asymmetric electron and hole concentration inside a device with $\mu_{pero}$ = 30 cm$^2$/Vs and $\mu_{ETL/HTL}$ = 10 cm$^2$/Vs on solar cell efficiency. (a) Defect recombination efficiency $\eta_R$ (s$^{-1}$) as a function of position inside the device and the energy level ($E_C$ - $E_T$) of the trap calculated for a flat-band device with symmetric ETL and HTL layer such that $n = p$ and is superimposed on the band diagram of the same device. (b) Same as before except now the ETL layer is doped at a donor concentration of $N_{d(ETL)} = N_C$. (c) Comparison of open circuit voltage V$_{oc}$ of (i) a device with symmetric contacts where $n$ and $p$ vary symmetrically through the device similar to that shown in Fig. 4c (represented by the green dashed line), (ii) a device with asymmetric contacts such that the ETL layer is doped with donor concentration of $N_{d(ETL)} = N_C$ (represented by the solid green curve) (ii) a flat band device with symmetric contacts such that $n = p$ (represented by the dashed red curve) and (iii) a device with asymmetric contacts such that the ETL layer is doped with donor concentration of $N_{d(ETL)} = N_c$ (represented by the solid red curve). (d) Solar cell efficiency of the same four devices mentioned in panel (c) is compared.

In Fig. 5c-d we compare $V_{oc}$ and $\eta$ (%) of the two flat-band devices. Also, to put their performance in perspective with that of device geometries discussed in Fig.4 we chose two devices, having contact layer mobility $\mu_{ETL/HTL}$ = 10 cm$^2$/Vs and absorber layer permittivity $\varepsilon_{pero}$ = 33.5. This makes sure that the electron and hole concentrations vary continuously through the device. One of the two devices with $\varepsilon_{pero}$ = 33.5 have symmetric contacts, i.e. undoped ETL or HTL while the ETL of the other device is doped with a donor concentration of $N_{d(ETL)} = N_{C(ETL)}$. It is observed from comparing the solid and dashed green curves representing the two devices with $\varepsilon_{pero}$ = 33.5 in both panel (c) and (d) that asymmetric contacts resulting in more asymmetric carrier concentration improve the $V_{oc}$ and $\eta$ (%) of midgap defects even when the n and p vary continuously inside the device. However, making the geometry such that n >> p across the device further inhibits the recombination of midgap defects even further. At midgap the maximum $V_{oc}$ and $\eta$ (%) are given by the solid red curves in Fig 5c and Fig 5d, respectively. The reference lines in Fig. 5d at 0.6eV, 0.8 eV and 0.9 eV away from the conduction band show that within 200 meV above midgap and 100 meV below midgap, the solar cell efficiency $\eta$ (%) improves as a function of the asymmetry between the electron and hole concentration inside the device.

However, even though n >> p helped reduce recombination through defects around and above midgap, it increases the recombination through defects 100 - 300 meV below midgap where $k_n$ << $k_p$. In a device where most defects are positioned such that $k_n$ << $k_p$, the HTL layer must be acceptor doped so that the electron concentration in the device remains small (n << p) and thus decreasing the electron capture rate $nk_n < pk_p$. Thus, the way to deal with asymmetric capture coefficients is to have asymmetric carrier concentration inside the device so that one of the capture rates is substantially slower than the other and $nk_n \neq pk_p$. However, to determine which geometry would improve the device efficiency, it is important to identify the dominant defect levels responsible for majority of the non-radiative recombination in the device by corroborating experimental data with first principle studies of capture coefficients. The knowledge of the capture coefficients of the dominant defect levels can then be used to choose the appropriate scheme of asymmetry in the device that successfully reduces the recombination efficiency of the dominant defect levels. In the following section we put this idea into practice and validate our hypothesis against real device data.

### C. Recombination through iodine interstitial.

Liu et. al[14] reported an open circuit voltage exceeding 1.26 V for inverted planar MAPI solar cells by carefully optimizing the hole transport layer and electron transport layer to an asymmetric champion device. In MAPI structures iodine interstitial defects have been long suspected to be the dominant defect level[4] and Zhang et.al[37] calculated the capture coefficients of iodine interstitial defects from first principles. Therefore, let us assume that iodine interstitials are the dominant defect in MAPI and use the capture coefficients calculated in ref. 37. Under these two assumptions we are able to make predictions on how the device geometry should affect device performance. We therefore simulate MAPI devices with an iodine interstitial defect level and varying asymmetries in the device geometry that modulate the relative efficiency of electron and hole injection.

We fabricated a MAPI solar cell of p-i-n structure with organic charge extraction layers poly(triarylamine) (PTAA) as the hole transport layer (p), [6,6]-phenyl-$C_{61}$-butyric acid methyl ester (PCBM) for the electron transport layer (n). The PTAA layer thickness is ~ 16 nm whereas the PCBM thickness is varied as ~ 10 nm, 25 nm, 40 nm, 55 nm and 63 nm. The current voltage curves are measured on a calibrated AM1.5 spectrum on a class AAA solar simulator providing a power density of 100 mW/cm$^2$. The $V_{oc}$ increases with increase in PCBM thickness due to suppression of recombination at interfaces and in the bulk leading to an improvement in $V_{oc}$ as shown in Fig . 6. Also the power output increased from 15.4 mW/cm$^2$ for the cell with $d_{ETL}$ = 10 nm to 20.4 mW/cm$^2$ for the cell with $d_{ETL}$ = 63nm.

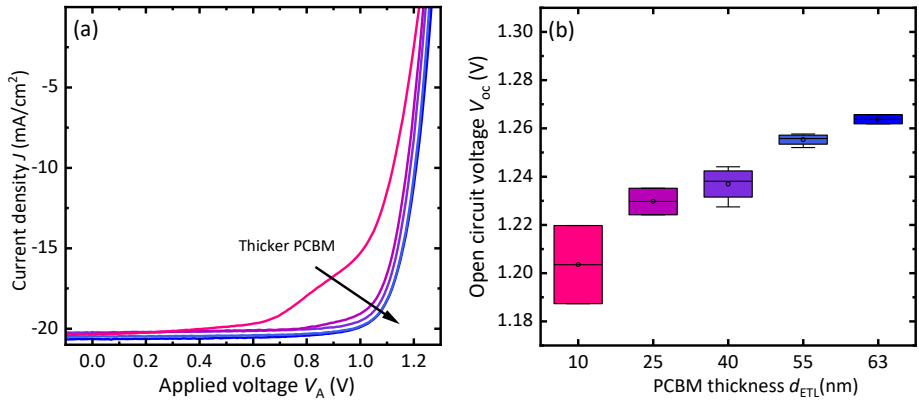

Fig 6(a) Experimental *JV* characteristics of inhouse fabricated standard MAPI cells showing the increase of open circuit voltage Voc with the thickness of the ETL. (b) Statistical data of open circuit voltage.

Recently Zhang et.al[37] performed first principle calculations to determine the electron and hole capture coefficients of Iodine interstitials defects about 0.48 eV away from the conduction band occurring in methylammonium lead iodide perovskite structures. Even though the iodine interstitial defect is energetically closer to the conduction band, they found the electron capture rate to be substantially slower as compared to the hole capture rate. The slowing down of the electron capture results from the strong anharmonicity in the potential energy surface of the system constituting of a positively charged iodine interstitial $I_i^+$, an e$^-$ at the conduction band minimum (CBM) and a h$^+$ at the valence band maximum (VBM) as shown in Fig.7. From a semiclassical perspective, this anharmonicity results in an energetic barrier $\Delta E_n^+$ between the potential energy surface PES of the two systems ($I_i^+ + e^- + h^+$ and $I_i^0 + h^+$) which an electron needs to overcome to transition from the CBM(green curve) to the defect level(orange curve). On the contrary, the electron "sees" no such barrier while transitioning from the defect level to the VBM(blue curve), thus making the hole capture rate substantially faster. To study the combined effects of asymmetric geometry and the asymmetric capture coefficients of an iodine interstitial defect on perovskite solar cell performance we take $k_n$ = 7 × 10$^{-9}$ cm$^3$/s and $k_p$ = 2 × 10$^{-5}$ cm$^3$/s as reported by Zhang et.al[37] from first principle calculations and simulate a symmetric and an asymmetric device for comparison.

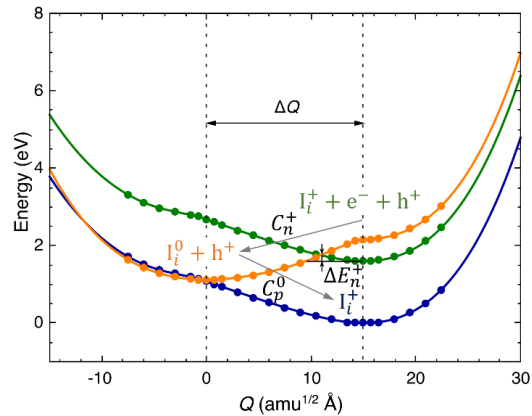

Fig. 7 The figure shows the potential energy surfaces for charge state transitions as a function of a generalized configuration coordinate. The green curve shows the potential energy surface of a system with an electron(e$^-$) at the conduction band minimum, a hole(h$^+$) at the valence band maximum and an iodine interstitial defect in its unoccupied state I$_i^+$. The I$_i^+$ defect captures an e$^-$ from the conduction band and transitions to the I$_i^0$ state represented by the orange curve. Semi-classically the process of transition from I$_i^+$ → I$_i^0$ needs to overcome a potential barrier of $\Delta E_n^+$ as determined by the intersection of the potential energy surfaces (green and the orange curve) of the two charge states. The defect I$_i^0$ then captures a hole from the valence band and relaxes the system back to I$_i^+$ represented by the blue curve. The small barrier of $\Delta E_n^+$ resulting from the strong anharmonicity of the potential energy surface of I$_i^+$ makes the electron capture process slower in comparison to the hole capture process as there is no such barrier between the orange and the blue curve. Reproduced with permission from American Physical Society. ©American Physical Society.[37]

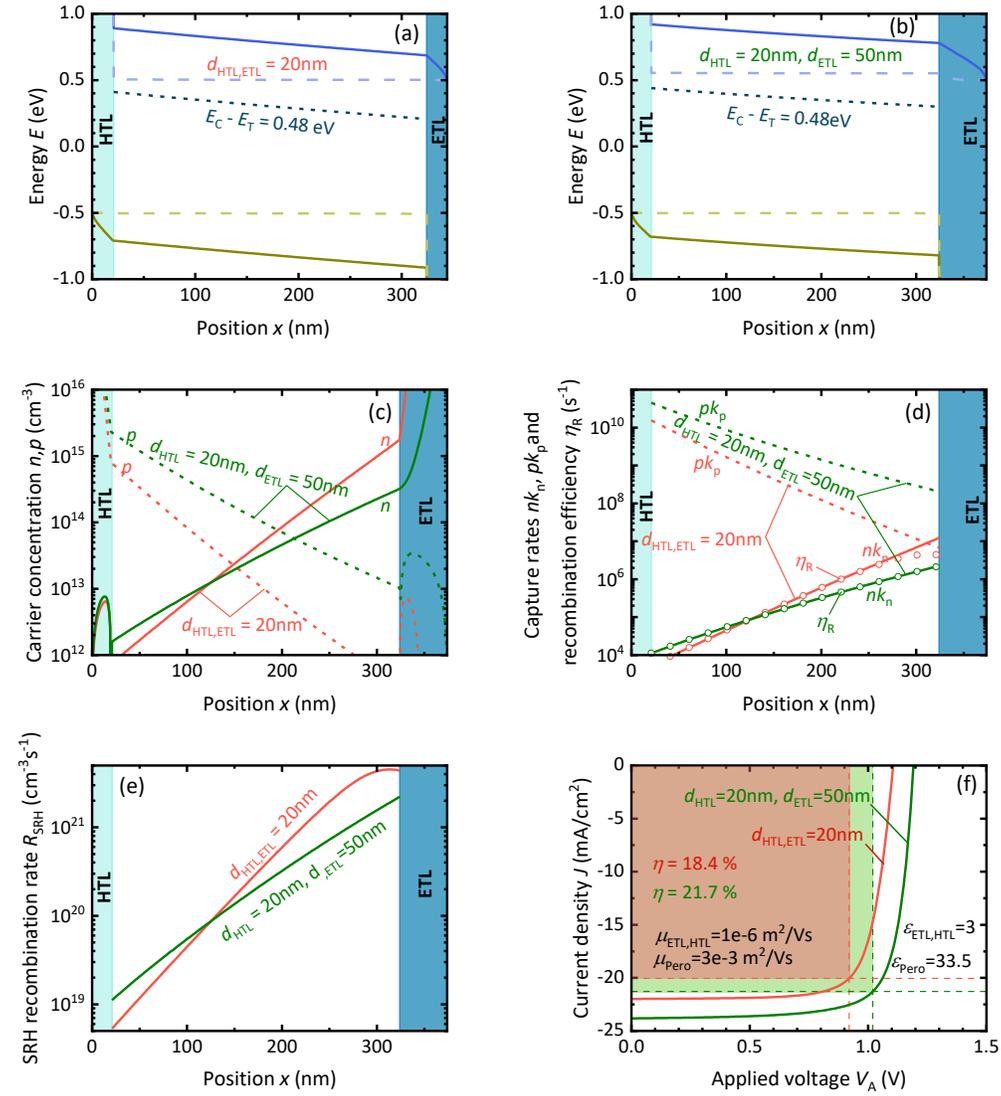

Fig. 8 Comparison of a symmetric and an asymmetric device with an iodine interstitial defect level. (a) The band diagram of the symmetric device with $d_{HTL,ETL}$ = 20 nm along with the iodine interstitial defect level 0.48 eV from the conduction band. (b) The same but for the asymmetric device which has $d_{HTL}$ = 20 nm and $d_{ETL}$ = 50 nm. (c) The electron concentration $n$ and hole concentration $p$ through the symmetric is represented by the curves in red and that through the asymmetric device is represented in green. The dashed curves represent the hole concentration and the solid curves the electron concentration. (d) The electron capture rate $nk_n$ (the solid curves), the hole capture rate $pk_p$ (the dashed curves), and the recombination efficiency $\eta_R$ (open symbols) through the absorber and the interfaces of the symmetric device are represented in red whereas that through the asymmetric device are represented in green. (e) The Shockley-Read-Hall recombination rate $R_{SRH}$ for a defect density of $N_T = 10^{15}$ / cm$^3$ across the absorber and the interfaces for the symmetric (red solid curve) and the asymmetric device (green solid curve). (f) The JV curve for the symmetric (red solid curve) and the asymmetric device (green solid curve).

In Fig. 8(a-b), we plot the band diagram of the symmetric and the asymmetric device, respectively, along with the iodine interstitial defect level 0.48 eV away from the conduction band. In the symmetric device, both the transport layers are identical with thickness $d_{HTL,ETL}$ = 20 nm with $\mu_{HTL,ETL}$ = 10$^{-2}$ cm$^2$/Vs whereas in the asymmetric device the ETL has a thickness of $d_{ETL}$ = 50 nm and HTL thickness is $d_{HTL}$ = 20 nm with the mobility remaining same as before. As a result of this asymmetry the electrostatic potential drop across the ETL increases and that across the absorber layer and the HTL decreases. The decrease in the potential drop in the absorber layer results in reduced electron concentrations inside the absorber and thus to maintain charge neutrality the hole concentrations in the absorber layer I increases as shown in Fig. 8(c). Now since the iodine interstitial defect has $k_n \ll k_p$, the smaller the electron concentration $n$ is compared to the hole concentration $p$ the better it is for the device as it makes $nk_n \ll pk_p$. This is shown in Fig. 8(d) where a comparison of the solid red and green curve reveals that the electron capture rate $nk_n$, which is the recombination limiting rate in this case, decreases with the introduction of the asymmetry. Even though the hole capture rate increases with increasing hole concentration, the "deep" defect criterion $nk_n = pk_p$ is completely avoided inside the asymmetric device. The recombination efficiency of the defect level in the asymmetric device represented by the green open symbols also decreases and is entirely determined by the electron capture rate $nk_n$. In Fig. 8(e) we obtain the non-radiative recombination rate $R_{SRH}$ through the iodine interstitial defect level for a defect density of $N_T$ = 10$^{15}$ / cm$^3$ across the interfaces and the absorber and we see a substantial decrease in the amount of non-radiative recombination. The reduced non-radiative recombination leads to an improvement in performance of the PSC as corroborated by the increase of $V_{oc}$ by approximately 100 meV and device efficiency $\eta$ (%) by more than 3% as shown in Fig. 8(f). Fig. 8(f) reproduces the same trend of increase in $V_{oc}$ with ETL thickness as studied from the experimental results in Fig. 6.

## D. Effect of the transport layer mobility on the device efficiency.

The carrier selective transport layers (TL) on either side of the perovskite absorber layer are responsible for the transport of the photogenerated charge carriers from the absorber layer to the extracting electrodes. However, the organic transport layers suffer from low mobilities such that $\mu_{HTL,ETL} \approx 10^{-5}$ cm$^2$/Vs – 10$^{-2}$ cm$^2$/Vs which is orders of magnitude lower than the mobility exhibited by the perovskite layer $\mu_{pero} \approx$ 1-10 cm$^2$/Vs[43]. These low mobilities of the state-of-the-art transport layers act as an efficiency limiting factor for PSC's. To investigate this relationship between TL mobility and device efficiency we choose to look at both symmetric as well as asymmetric perovskite device structures with a single defect level given by the iodine interstitial defect as was used in the previous section. In Fig. 9(a-d) we plot the band diagrams of four different device geometries at three different values of TL mobilities such that $\mu_{HTL,ETL}$ = [$10^{-5}$ cm$^2$/Vs, $10^{-2}$ cm$^2$/Vs, 10 cm$^2$/Vs] to show the position of the quasi-Fermi level splitting as a function of the transport layer mobilities.

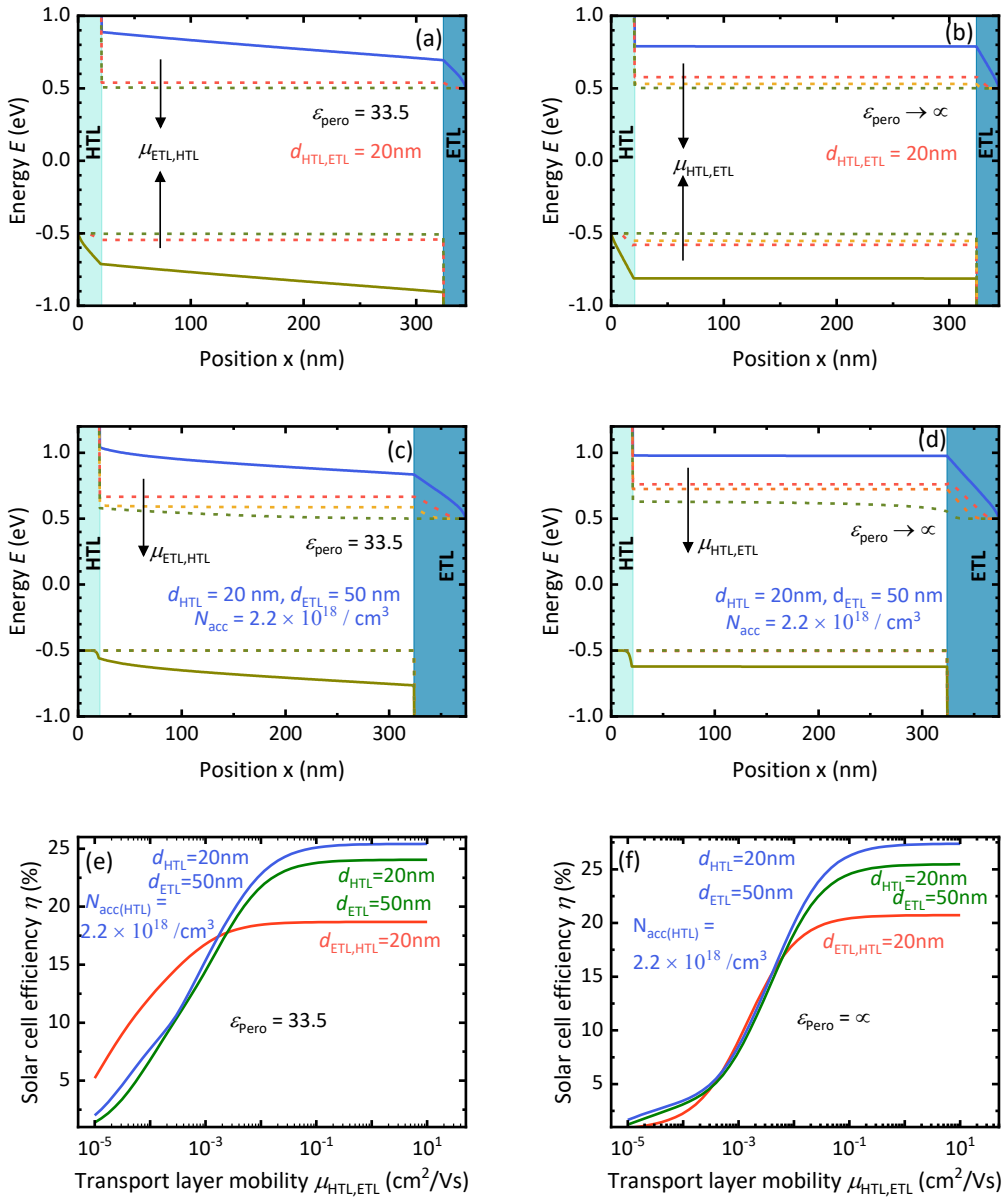

Fig. 9 Effect of transport layer mobility on the efficiency of the device. (a) Band diagram of a symmetric device with $\varepsilon_{pero} \rightarrow 33.5$, $d_{HTL,ETL}$ = 20nm and with one defect level at 0.48 eV away from the conduction band at three different values of transport layer mobility $\mu_{HTL,ETL}$ = [$10^{-5}$ cm$^2$/Vs, $10^{-2}$ cm$^2$/Vs, 10 cm$^2$/Vs]. (b) Same as in (a) but now the permittivity of the absorber $\varepsilon_{pero} \rightarrow \infty$. (c) Band diagram of an asymmetric device with $\varepsilon_{pero} \rightarrow 33.5$, $d_{ETL}$ = 50 nm, $d_{HTL}$ = 20nm, acceptor doped HTL layer and with one defect level at 0.48 eV away from the conduction band at three different values of transport layer mobility as before. (d) Same as in (c) but now the permittivity of the absorber $\varepsilon_{pero} \rightarrow \infty$. (e) The efficiency $\eta$ (%) of the different solar cell geometries with $\varepsilon_{pero} \rightarrow 33.5$ as a function of transport layer mobility. (f) Same as in (e) but for solar cell geometries with $\varepsilon_{pero} \rightarrow \infty$.

In case of the undoped device geometries as shown in Fig. 9(a-b) for $\varepsilon_{pero}$ = 33.5 and $\varepsilon_{pero} \rightarrow \infty$, respectively, increasing the TL mobility from $10^{-5}$ cm$^2$/Vs to 10 cm$^2$/Vs moves both the electron quasi-Fermi level ($E_{fn}$) as well as the hole quasi-Fermi ($E_{fp}$) level away from the conduction band and valence band, respectively. As a result of the faster transport of carriers through the TL's, electron and hole concentration inside the device decreases thereby inhibiting recombination. This enhances the efficiency of the device as shown by the red curve in Fig. 9(e-f) for $\varepsilon_{pero}$ = 33.5 and $\varepsilon_{pero} \rightarrow \infty$, respectively. It is important to note the sensitivity of the recombination and hence the device efficiency on the carrier concentration inside the device. In Fig. 9a, the $E_{fn}$ and $E_{fp}$ for $\mu_{HTL,ETL}$ = $10^{-2}$ cm$^2$/Vs and $\mu_{HTL,ETL}$ = 10 cm$^2$/Vs overlaps and this in turn makes the efficiency of the device almost constant when the TL mobility varies between $\mu_{HTL,ETL}$ = $10^{-2}$ cm$^2$/Vs and 10 cm$^2$/Vs as shown in Fig. 9e.

In Fig. 9(c-d) the asymmetric device has a longer ETL and a doped HTL. As a result of the doping of the HTL, the $E_{fp}$ is pinned to the valence band and does not move away from the valence band even when the TL mobility is increased. However, the $E_{fn}$ moves away from the conduction band and we see a significant improvement in the efficiency of the device as shown by the blue curve in Fig. 9(e-f). This is because the recombination through the defect level, which in this case is the iodine interstitial level, is limited by the slower of the two capture rates. The decrease of the electron concentration inside the device with the increase of TL mobility reduces the electron capture rate within the device and thus improving the device efficiency.

When the TL mobilities $\mu_{HTL,ETL}$ < $10^{-2}$ cm$^2$/Vs and are thus much slower than the perovskite mobility, even introducing the asymmetry in the device would not necessarily improve the device efficiency as seen in both Fig. 9(e-f). The poor TL mobility results in slow charge transport leading to charge accumulation inside the transport layers as denoted by the bending of quasi-Fermi level splitting along the curvature of the conduction band or valence band of the TL's. As an effect of the bending of the quasi-Fermi level along the bands, the carrier concentration inside the absorber layer is much higher and the efficiency lower.

## IV. Conclusion

Defects play a key role in limiting conversions efficiencies of photon energy to electrical energy by enabling alternate channels of carrier recombination for excited carriers to escape from their respective bands before they could be extracted at the contacts. The recombination rates depend on the actual values of the capture coefficients for electrons and holes and on their densities. Because capture of electrons and holes are two processes that have to happen in series to allow for a single recombination event, it is always the slower of the two processes that limits the time constant for recombination. The capture rates depend on the relative concentrations of electrons and holes and thereby also on the device geometry and the electrostatics of the solar cell. A device geometry that maximizes the volume where one capture rate is particularly slow and therefore limits the total recombination rate would therefore be one that reduces the relative impact of defects on device performance. It would be a defect tolerant device geometry.

The answer to the question which device geometry is best does, however, not have a generic answer. We have to distinguish between different situations. In particular, we have to consider the ratio of the capture coefficients, i.e. how easy or difficult is it for a certain defect to capture one carrier relative to capturing the other. Defect assisted recombination is often modelled using equal capture coefficients for electrons and holes, however, we show here that this is in general a rather unlikely scenario. Evidence for this claim comes both from calculations using the harmonic approximation but also from literature data of actual defects in halide perovskites. With the assumption of asymmetric capture coefficients of a defect, we can modify a device geometry in such a way that the density of the carrier associated with the slower of the two capture coefficients decreases inside the absorber layer. This decrease will then slow down recombination. The implementation of these generic design principles can be done in a variety of ways and will depend on the device and the dominant defect levels in question.

After discussing the general principles of defect tolerant deice geometries, we focus on one relevant example of recombination in methylammonium lead halide based solar cells. We base our calculations on assuming that the capture coefficients calculate by Zhang et al. for iodine interstitials are correct and dominating recombination in the lead-halide absorber layer. Since Zhang predict higher capture coefficients for holes than for electrons, the recombination could be further reduced by slowing down the slower of the two rates (the electron capture) by reducing the electron concentration and increasing the hole concentration. This can be done in various ways with p-type doping being the most obvious approach from a conceptual point of view. Due to their ionic nature, the creation of stable doping profiles in halide perovskites is however challenging. Thus, we show how changing the thickness of charge transport layers can be used to modulate the electron and hole concentrations and thereby improve photovoltaic performance. We find that slightly higher thicknesses for the ETL than for the HTL should be beneficial for device performance. This interestingly agrees with experimental findings in MAPI solar cells with high open-circuit voltages.

## V. Experimental Section

### A. Device Fabrication

1. Materials.
Methylammonium iodide (MAI) was purchased from Greatcell Solar. Lead acetate trihydtrate (Pb(CH$_3$COO)$_2$·3H$_2$O, >99.5%) and Bathocuproine (BCP, >99.8%) were purchased from TCI. Lead chloride (PbCl$_2$, >99.999%) was purchased from Sigma-Aldrich Company Ltd. Poly[bis (4-phenyl)(2,4,6-trimethylphenyl)amine] (PTAA, Mn = 17900, Mw = 33000) was purchased from purchased from Xi'an Polymer Light Technology Corp (China). [6,6]-phenyl-C$_{61}$-butyric acid methyl ester (PCBM) was purchased from Solenne (Netherlands). Toluene (T), purity of 99.8%), *N,N*-dimethylformamide (DMF, 99.8%), isopropanol (IPA, 99.5%) and chlorobenzene (CB, 99.8%) were purchased from Sigma-Aldrich and used as received.

2. Device fabrication.

The pre-patterned ITO substrates (2.0 × 2.0 cm$^2$) were bought from KINTEC (Hong Kong) and ultrasonically cleaned with soap (Hellmanex III), deionized water, acetone and IPA in succession for 10 min. The as-cleaned ITO substrates were treated with oxygen plasma for 12 min and transferred to a N$_2$-filled glovebox. 80 μl PTAA (2 mg ml$^{-1}$ in toluene) solution was spin-coated onto the ITO substrates with a two-consecutive step program at 500 rpm for 4 s (with a ramping rate of 500 rpm s$^{-1}$) and 4500 rpm for 20 s (with a ramping rate of 800 rpm s$^{-1}$), then the samples were thermally annealed at 100 °C for 10 min and afterwards cooled down to room temperature. The PTAA layer thickness is ~16 nm. The perovskite precursor solution prepared by mixing Pb(CH$_3$COO)$_2$·3H$_2$O (0.54 M), PbCl$_2$ (0.06 M), DMSO (0.9 M) and MAI (1.8 M) in DMF was stirred at room temperature for 60 min and filtered with a 0.45 μm PTFE filter prior to use. To fabricate the perovskite layer, 120 μl perovskite precursor solution was spin-coated on the top of PTAA layer by a two-consecutive step program at 1400 rpm for 15 s (with a ramping rate of 350 rpm s$^{-1}$) and 6000 rpm for 40 s with a ramping rate of 767 rpm s$^{-1}$. The samples were immediately annealed on a hotplate at 75 °C for 2 min. Afterwards they were cooled down to room temperature. 60 ul PCBM solution (5, 10, 15, 20 and 25 mg ml$^{-1}$ in CB and toluene) was spin-coated on the top of perovskite layer at a speed of 1200 rpm for 60 s (with a ramping rate of 400 rmp s$^{-1}$) as electron transport layer (ETL). The thickness is 10nm, 25 nm, 40 nm, 55 nm respectively. For the drying of the PCBM layer the samples were left in an open petri dish for 20 min, without additional annealing. Then the samples were spin-coated 100 μl BCP (0.5 mg ml$^{-1}$ in IPA) at 4000 rpm for 30 s (with a ramping rate of 800 rpm s$^{-1}$). Finally, 8 nm BCP and 80 nm Ag was thermally evaporated in a separate vacuum chamber (<5×10$^{-6}$ Pa) through a metal shadow mask to define an aperture area of 0.16 cm$^2$ by the overlap of the ITO and the Ag.

3. Device characterization.

*Current-voltage-characterization (JV)* :

The current-voltage curves were measured on a calibrated AM1.5 spectrum of a class AAA solar simulator (WACOM-WXS-140S-Super-L2 with a combined xenon/ halogen lamp-based system) providing a power density of 100 mW/cm$^2$. The forward scan (-0.1 V to 1.27 V) and the subsequent reverse voltage scan (1.27 V to -0.1 V) were each carried out at a scan speed of 100 mV/s, using a Series 2420 SourceMeter (Keithley Instruments). All measurements were carried out under inert atmosphere in a sealed, electrically contacted measurement box in glovebox. Each sample contains four solar cells with an active cell area of 0.16 cm$^2$.

*LED-solar simulator measurements*:

In addition to the current-voltage characterization on the calibrated class AAA solar simulator, another setup directly integrated into the glove box was used for (JV-curves) and maximum power point (MPP) and $V_{oc}$-tracking. This solar simulator is equipped with a white light LED (Cree XLamp CXA3050), whose illuminance has been adjusted to one sun conditions using the short-circuit current resulting from the EQE measurement of a perovskite cell. A 2450 Keithley is used as a source measure unit.

# Supplementary information for Defect tolerant device geometries.

Basita Das[1], Irene Aguilera[1], Uwe Rau[1] and Thomas Kirchartz[1,2]
[1]IEK5-Photovoltaik, Forschungszentrum Jülich, 52425 Jülich, Germany
[2]Faculty of Engineering and CENIDE, University of Duisburg-Essen, Carl-Benz-Str. 199, 47057 Duisburg, Germany

1. Analytical expressions

Table I Generalized microscopic model for calculating multiphonon capture coefficients of defects.

| Expressions for quantum defect model that describes the connection between depth of a defect and the radius of the defect wavefunction[44,45] | |
|---|---|
| Quantum defect parameter ($\upsilon_\text{T}$) | $\upsilon_\text{T} = \sqrt{R_{\text{H}^*}/\Delta E_{\min}} = \dfrac{1}{\epsilon_\infty}\sqrt{\dfrac{m^* q^4}{32\pi^2 \Delta E_{\min}}}$ |
| Radius of the deep defect wavefunction ($R_\text{T}$) | $R_\text{T} = \dfrac{a_\text{H}^* \upsilon_\text{T}}{2}$ |
| **Expressions for calculation of non-radiative multiphonon capture coefficients[39,46,47]** | |
| Non-radiative multiphonon capture coefficient | $k_{\text{n/p}} = \dfrac{M_{\text{i,f}}^2 \sqrt{2\pi}}{\hbar^2 \omega_\nu \sqrt{l\sqrt{1+x^2}}} \exp\left[l\left(\dfrac{\hbar\omega_\nu}{2 k_\text{B} T} + \sqrt{1+x^2}\right.\right.$ $\left.\left. - x \cosh^{-1}\left(\dfrac{\hbar\omega_\nu}{2k_\text{B}T}\right) - \ln\left(\dfrac{1+\sqrt{1+x^2}}{x}\right)\right)\right]$ |
| No. of phonons emitted during multiphonon transition | $l = \dfrac{\Delta E}{\hbar\omega_\nu}$ |
| Multiphonon transition matrix element | $\lvert M_{\text{i,f}}\rvert^2 = V_\text{T}(l\hbar\omega_\nu)^2$ |
| Volume of the defect $V_\text{T}$ | $V_\text{T} = \dfrac{4}{3}\pi R_T^3$ |
| Parameter $x$ | $x = \begin{cases} \dfrac{S_\text{HR}}{l \sinh(\hbar\omega_\nu/2k_\text{B}T)} & \text{for}\quad S_\text{HR} < l \\ \dfrac{l}{S_\text{HR}\sinh(\hbar\omega_\nu/2k_\text{B}T)} & \text{for}\quad S_\text{HR} > l \end{cases}$ |
| Sommerfeld factor | $s_\text{a} = 4(\pi R_\text{H}^*/k_B T)^{1/2}$ |
| **Expressions for calculation of Huang-Rhys factor[40]** | |
| Huang-Rhys factor for polar optical coupling | $S_\text{HR} = \dfrac{3}{2(\hbar\omega_\nu)^2}\left[\dfrac{q^2\,\hbar\omega_\nu}{a_0^3\,q_\text{D}^2}\left(\dfrac{1}{\epsilon_\infty} - \dfrac{1}{\epsilon}\right)\right] I\left(-2,2\mu,\dfrac{q_\text{D} a_\text{H}^* \upsilon_\text{T}}{2}\right)$ |
| Huang-Rhys factor for optical deformation potential coupling | $S_\text{HR} = \dfrac{1}{2(\hbar\omega_\nu)^2}\dfrac{\hbar D^2}{M_\text{r}\omega_\nu} I\left(0,2\mu,\dfrac{q_\text{D} a_\text{H}^* \upsilon_\text{T}}{2}\right)$ |

| Function I | $I(a, b, c) = \dfrac{1}{(bc)^2} \displaystyle\int_0^1 \dfrac{y^a \sin^2(b \tan^{-1}(cy))}{[1+(cy)^2]^b} dy$ |
|---|---|
| Additional expressions | |
| Radius of the sphere of the Brillouin zone $q_D$ | $q_D = \sqrt[3]{6\pi^2}/a_0$ |
| Bohr radius $a_H$ | $a_H = 4\pi\epsilon_0/mq^2$ |
| Effective Bohr radius $a_H^*$ | $a_H^* = a_H \epsilon/m^*$ |
| Rydberg energy $R_H$ | $R_H = q^2/(8\pi\epsilon_0 a_H)$ |
| Effective Rydberg energy $R_H^*$ | $R_H^* = q^2/(8\pi\epsilon a_H^*)$ |

Table II Material parameters used in simulation

| | |
|---|---|
| Thickness of Absorber, $d_{pero}$ | 300 nm |
| Thickness of Hole transport layer, $d_{HTL}$ | 20 nm |
| Thickness of Electron transport layer, $d_{ETL}$ | 20 nm (Fig. 4, 5 and 6a) |
| | 50 nm (Fig. 6b, 7b and 7c) |
| Thickness of interfaces (HTL/Pero, ETL/Pero) | 2 nm |
| Electron affinity of Absorber, $E_{A(Pero)}$ | 4 eV |
| Electron affinity of HTL, $E_{A(HTL)}$ | 2.6 eV |
| Electron affinity of ETL, $E_{A(Pero)}$ | 4 eV |
| Bandgap of Absorber, $E_{g(Pero)}$ | 1.6 eV |
| Bandgap of HTL, $E_{g(HTL)}$ | 3 eV |
| Bandgap of ETL, $E_{g(ETL)}$ | 4 eV |
| Bandgap of HTL/Pero interface | $E_{g(HTL)} + E_{A(HTL)} - E_{A(Pero)}$ |
| Bandgap of ETL/Pero interface | $E_{g(Pero)} + E_{A(Pero)} - E_{A(ETL)}$ |
| Mobility of Absorber, $\mu_{(Pero)}$[43] | 30 cm$^2$/Vs |
| Mobility of HTL, ETL $\mu_{(HTL,ETL)}$[24] | 10$^{-2}$ cm$^2$/Vs Fig. 4, Fig. 6 |
| | 10 cm$^2$/Vs Fig. 5 |
| | 10$^{-5}$ – 10 cm$^2$/Vs Fig. 7 |
| Effective density of carriers (all layers) | 2.2 × 10$^{18}$ cm$^{-3}$ |
| Direct recombination coefficient (all layers) | 5 × 10$^{-11}$ cm$^3$/s |
| Density of Donor traps (Interface and absorber layers) | 10$^{15}$ cm$^{-3}$ |
| Electron capture coefficient ($k_n$)[30,37] | Variable (Fig. 4 and 5) |
| | 0.7 × 10$^{-8}$ cm3/s (Fig. 6 and 7) |
| Hole capture coefficient ($k_p$) | Variable (Fig. 4 and 5) |
| | 0.2 × 10$^{-4}$ cm3/s (Fig. 6 and 7) |

2. Surface recombination limited devices.

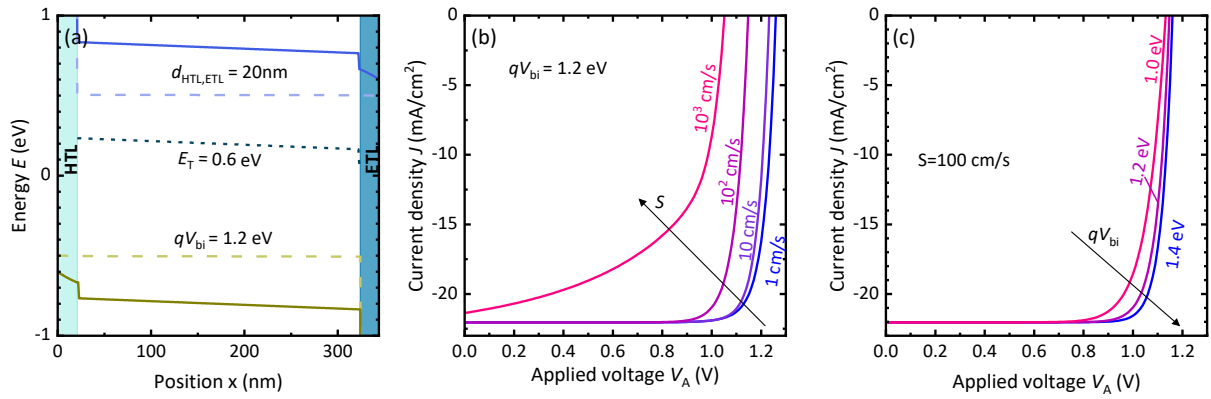

Fig S10 Interface limited recombination in devices. (a) Band diagram of a symmetric device with a conduction band offset and valence band offset of 0.1eV plotted at $V_A$ = 1.0V . (b) JV plots of the device at fixed qVbi = 1.2 eV and varying surface recombination velocity $S$. The device performance deteriorates with increase in $S$. (c) JV plots of the same device a fixed surface recombination velocity S = 100 cm/s and varying qVbi. The trends in device performance are similar to that observed in Fig 4 for decreasing qVbi.